\documentclass{aa}

\begin{document}

   \title{Automated Clustering Algorithms for Classification of Astronomical Objects}
   \author{Yanxia Zhang,$^{1}$
           Yongheng Zhao$^{1}$
                      } 

   \offprints{Yanxia Zhang}

   \institute {\inst{1}National Astronomical Observatories, Chinese Academy of Sciences, Beijing 100012, P.R. China.\\
                       email: zyx@lamost.bao.ac.cn, yzhao@lamost.bao.ac.cn \\
              }

    \date{Received ; accepted }
\authorrunning{Zhang Yanxia, Zhao Yongheng}
\maketitle
   
\begin{abstract}
Data mining is an important and challenging problem for the efficient analysis of large astronomical databases and will become even more important with the development of the Global Virtual Observatory. In this study, learning vector quantization (LVQ), single-layer perceptron (SLP) and support vector machines (SVM) were put forward for multi-wavelength data classification. A feature selection technique was used to evaluate the significance of the considered features to the results of classification. From the results, we conclude that in the situation of less features, LVQ and SLP show better performance. In contrast, SVM shows better performance when considering more features. The focus of the automatic classification is on the development of efficient feature-based classifier. The classifiers trained by these methods can be used for preselecting AGN candidates.


\end{abstract}

\section{Introduction}

Today, there are many impressive archives painstakingly constructed from observations associated with an instrument. The Hubble Space Telescope (HST), the Chandra X-Ray Observatory, the Sloan Digital Sky Survey (SDSS), the Two Micron All Sky Survey (2MASS), and the Digitized Palomar Observatory Sky Survey (DPOSS) are examples of this. Furthermore yearly advances in electronics bring new instruments, doubling the amount of data we collect each year. For example, approximately a gigapixels is deployed on all telescopes today, and new gigapixel instruments are under construction. This trend is bound to continue. Just like what Szalay says, the astronomy is facing "data avalanche" (See e.g., Szalay \& Gray 2001). How to organize, use, and make sense of the enormous amounts of data generated by today's instruments and experiments? It is very time consuming and demands high quality human resources. Therefore, better features and better classifiers are required. In addition, expert systems are also useful to get quantitative information. 

It is possible to solve the above questions with Neural Networks (NNs), because they permit application of expert knowledge and experience through network training. Furthermore, astronomical object classification based on neural networks requires no priori assumptions or knowledge of the data to be classified as some conventional methods need. Neural networks , over the years, have proven to be a powerful tool capable to extract reliable information and patterns from large amounts of data even in the absence of models describing the data (cf. Bishop 1995) and are finding a wide range of applications also in the astronomical community: catalogue extraction (Andreon et al. 2000), star/galaxy classification (Odewahn et al. 1992; Naim et al. 1995; Miller \& Coe 1996; M$\ddot {a}$h$\ddot {o}$nen \& Hakala 1995; Bertin \& Arnout 1996; Bazell \& Peng 1998), galaxy morphology (Storrie-Lombardi et al. 1992; Lahav et al. 1996), classification of stellar spectra (Bailer-Jones et al. 1998; Allende Prieto et al. 2000; Weaver  2000). Just to name a few, the rising importance of artificial neural networks is confirmed in this kind of task. There is also a very important and promising recent contribution by Andreon et al. (2000) covering a large number of neural algorithms.

In this work, a class of supervised neural networks called learning vector quantization (LVQ) was proposed. LVQ shares the same network architecture as the Kohonen self-organizing map (SOM), although it uses a supervised learning algorithm. Bazell \& Peng (1998) pioneered the use of it in astronomical applications. Another class of supervised neural networks named multi-layer perceptrons (MLP) was presented. Goderya \& McGuire (2000) summarized progress made in the development of automated galaxy classifiers using neural networks including MLP. Qu et al. (2003) experimented and compared multi-layer perceptrons (MLP), radial basis function (RBF), and support vector machines (SVM) classifiers for solar-flare detection. Meanwhile, an automated algorithm called support vector machines (SVM) for classification was introduced. The approach was originally developed by Vapnik (1995). Wozniak et al. (2001) and Humphreys et al. (2001) have pioneered the use of SVM in astronomy. Wozniak et al. (2001) evaluated SVM, K-means and Autoclass for automated classification of variable stars and compared their effectiveness. Their results suggested a very high efficiency of SVM in isolating a few best defined classes against the rest of the sample, and good accuracy for all classes considered simultaneously. Humphreys et al. (2001) used different classification algorithms including decision trees, K-nearest neighbor and support vector machines for classifying the morphological type of the galaxy. Furthermore, they got the very promising results of their first experiments with different algorithms.

Celestial objects radiate energy over an extremely wide range of wavelengths from radio waves to infrared, optical to ultraviolet, X-ray and even gamma rays. Each of these observations carries important information about the nature of objects. Different physical processes show different properties in different bands. Based on these, we apply learning vector quantization (LVQ), single-layer perceptron (SLP) and support vector machines (SVM) to classify AGNs, stars and normal galaxies with data from optical, X-ray, infrared bands. In this paper we present the principles of LVQ, SLP and SVM in section 2. In section 3, we discuss the sample selection and analysis the distribution of parameters. In section 4 the computed results and discussion are given. Finally, in section 5 we conclude this paper with a discussion of general technique and its applicability.

\section{The Methods Used}

\subsection{Learning Vector Quantization}

Here the adopted learning vector quantization (LVQ) algorithm is based upon the LVQ\_PAK routines developed at the Laboratory of Computer and Information Sciences, Helsinki University of Technology, Finland. Their software can be obtained via the WWW from www.cis.hut.fi/research/lvq\_pak/. If interested in the application of LVQ in astronomy, we can refer to the papers of Bazell \& Peng (1998) and Cortiglioni et al. (2001).

The LVQ method was developed by Kohonen (1989) who also developed the popular unsupervised classification technique known as the self-organizing map or topological map neural networks (Kohonen 1989, 1990). SOM performs a mapping from an $n$-dimensional input vector onto two-dimensional array of nodes that is usually displayed in a rectangular or hexagonal lattice. The mapping is performed in such a way as to preserve the topology of the input data. This means that input vectors that are similar to each other in some sense, are mapped to neighboring regions of the two-dimensional output lattice. Each node in the output lattice has an $n$-dimensional reference vector of weights associated with it, one weight for each element of the input vector. The SOM functions compare the distance, in some suitable form, between each input vector and each reference vector in an iterative manner. With each iteration, the reference vectors are moved around in the output space until their positions converge to a stable state. When the reference vector that is closest to a given input vector is found (the winning reference vector), that reference vector is updated to more closely match the input vector. This is the learning step.

LVQ uses that same internal architecture as SOM: a set of $n$-dimensional input vectors are mapped onto a two-dimensional lattice, and each node on the lattice has an $n$-dimensional reference vector associated with it. The learning algorithm for LVQ, i.e., the method of updating the reference vectors, is different from that of SOM. Because LVQ is a supervised method, during the learning phase the input data are tagged with their correct class. We define the input vector $x$ as:
\begin{equation}
x=(x_1,x_2,x_3,\cdots,x_n) 
\end{equation} 
Reference vector for $i$th output neuron ${\omega}_{i}$ as:
\begin{equation}
{\omega}_{i}=( {\omega}_{1i},{\omega}_{2i},{\omega}_{3i},\cdots,{\omega}_{ni})
\end{equation} 
Define Euclidean distance between the input vector and the reference vector of
the neuron $i$ as:
\begin{equation}
D(i)=\sqrt{\sum_{j=1}^{n}(x_j-{\omega}_{ji})^2}
\end{equation}
When $D(i)$ is a minimum, the input vectors are compared to the reference
vectors and the closest match is found using the formula
\begin{equation}
\mid {\omega}_{i^*}-x \mid \le \mid {\omega}_{i}-x\mid,
\end{equation}
 where $x$ is an input vector, ${\omega}_{i}$ are the reference vectors, and ${\omega}_{i^*}$ is the winning reference vector. The reference vectors are then updated using the following rules:\\
If $x$ is in the same class as ${\omega}_{i^*}$,
\begin{equation} \Delta {\omega}_{i^*}= \alpha (t)(x-{\omega}_{i^*})
\end{equation}
If $x$ is in a different class from ${\omega}_{i^*}$,
 \begin{equation}
 \Delta{\omega}_{i^*}= -\alpha (t)(x-{\omega}_{i^*}) \qquad
\end{equation}
If $i$ is not the index of the winning reference vector,
 \begin{equation}
\Delta {\omega}_{i^*}= 0 \qquad
\end{equation}

The learning rate $0<\alpha (t)<1$ should generally be made to decrease monotonically with time, yielding larger changes for early iterations and more fine tuning as convergence is approached. The time $t$ is taken as positive integers.  Here we adopt the optimized-leaning-rate $\alpha (t)$ (see Kohonen et al. 1995)
\begin{equation}
\alpha (t)={\alpha (t-1)\over 1+s(t)\alpha (t-1)}
\end{equation}
where $s(t)=+1$ if the classification is correct and $s(t)=-1$ if the classification is wrong. In this work, the initial value of $\alpha (t)$ is selected, 0.3, whereby learning is significantly speeded up, especially in the beginning, and the ${\omega}_{i^*}$ quickly find their approximate asymptotic values. Two hundred codebook vectors in the codebook is adopted, meanwhile, 7 neighbors is used in knn-classification. The network is trained for 5000 epochs. There are several versions of the LVQ algorithm for which the learning rules differ in some details. See Kohonen (1995) for an explanation of the differences between these algorithms. When the learning phase is over, the reference vectors can be frozen, and any further inputs to the system will be placed into one of the existing classes, but the classes will not change.

\subsection{Support Vector Machines}
Support Vector Machines (SVM) are learning machines that can perform binary classification (pattern recognition) and real valued function approximation (regression estimation) tasks. SVM creates functions from a set of labeled training data and operate by finding a hypersurface in the space of possible inputs. This hypersurface will attempt to split the positive examples from the negative examples. The split will be chosen to have the largest distance from the hypersurface to the nearest of the positive and negative examples. Intuitively, this makes the classification correct for testing data that is near, but not identical to the training data. In detail, during the training phase SVM takes a data matrix as input, and labels each sample as either belonging to a given class (positive) or not (negative). SVM treats each sample in the matrix as a point in a high-dimensional feature space, where the number of attributes determines the dimensionality of the space. SVM learning algorithm then identifies a hyperplane in this space that best separates the positive and negative training samples. The trained SVM can then be used to make predictions about a test sample's membership in the class. In brief, SVM non-linearly maps their n-dimensional input space into a high dimensional feature space. In this high dimesional feature space a linear classifier is constructed. More information can be found in Burges' tutorial (1998) or in Vapnik's book (1995).

Given some training data\\
$$
 (x_1,y_1),...,(x_l,y_l),\qquad y_i\in {(-1,1)}
$$

If the data is linearly separable, one can separate it by an infinite number of linear hyperplanes. We can write these hyperplanes as
\begin{equation}
f(x,\alpha)=({\omega}_{\alpha} \cdot x)+b
\end{equation}

Among these hyperplanes, the one with the maximum margin is called by the optimal separating hyperplane. This hyperplane is uniquely determined by the support vectors on the margin. It satisfies the conditions  
\begin{equation}
y_i[(\omega \cdot x_i)+b]\ge 1,\qquad i=1,\ldots,l.
\end{equation}

Besides satisfying the above conditions, the optimal hyperplane has the minimal norm
\begin{equation}
||\omega||^2=(\omega \cdot \omega)
\end{equation}

The optimal hyperplane can be found by finding the saddle point of the Lagrange functional:
\begin{equation}
L(\omega,b,\alpha)=\frac{1}{2}\omega \cdot \omega -\sum_{i=1}^{l}
{\alpha}_i([(\omega \cdot x_i)+b]y_i-1)
\end{equation}
where ${\alpha}_i$ are Lagrange multipliers. The Lagrangian has to be minimized with respect to $\omega$,$b$ and maximized with respect to ${\alpha}_i \ge 0$.

The saddle point is defined as follows:
\begin{equation}
\omega=\sum_{i=1}^{l} {\alpha}_{i}x_{i}y_{i}
\end{equation}
where $\alpha$ is the maximum point of 
\begin{equation}
W(\alpha)=\sum_{i=1}^{l}{\alpha}_{i}-\frac{1}{2}\sum_{i,j=1}^{l}{\alpha}_{i}{\alpha}_{j}y_iy_j(x_i \cdot x_j)
\end{equation}
subject to constraints
\begin{equation}
\sum_{i=1}^{l}{\alpha}_{i}y_i=0, \qquad {\alpha}_i \ge 0 
\end{equation}

Therefore the optimal separating hyperplane has the form
\begin{equation}
f(x)=sign(\sum_{support \,\, vectors}y_i{\alpha}_{i}(x_i \cdot
x)-b)
\end{equation}

This solution only holds for linearly separable data, but has to be slightly modified for linearly non-separable data, the
${\alpha}_{i}$ has to be bounded:
\begin{equation}
0\le{\alpha}_{i} \le C
\end{equation}
where C is a constant chosen a priori.

To generalize to non-linear classification, we replace the dot product with a kernel [$k(\cdot)$]. For binary classification, Stitson et al.(1996) and Gunn (1998) stated it in detail. As for the multi-class classification can refer to Weston and Watkins (1998).

\subsection{Single-layer Perceptron} 

Multi-layer perceptrons (MLP) are feedforward neural networks trained with the standard backpropagation algorithm. If no hidden layer, MLP are also called single-layer perceptron. They are supervised networks so they require a desired response to be trained. They learn how to transform input data into a desired response, so they are widely used for pattern classification. With one or two hidden layers, they can approximate virtually any input-output map. They have been shown to approximate the performance of optimal statistical classifiers in difficult problems. Most neural network applications involve MLP.

The basic MLP building unit is a model of artificial neuron. This unit computes the weighted sum of the inputs plus the threshold weight and passes this sum through the activation function (usually sigmoid) (18), (19):
\begin{equation}
\nu_j=\theta_j+\sum_{i=1}^p\omega_{ji}x_i
\end{equation}
\begin{equation}
y_j=\varphi(\nu_j)
\end{equation}
where $\nu_j$ is the linear combination of inputs $x_1,x_2,\ldots,x_p$ of neuron $j$, $\omega_{j0}=\theta_j$ is the threshold weight connected to a special input $x_0=-1$, $y_j$ is the output of neuron $j$, and $\varphi_j(\cdot)$ is its activation function. Herein we use a special form of sigmoidal (non-constant, bounded, and monotone-increasing) activation function - logistic function
\begin{equation}
y_j=\frac{1}{1+exp(-\nu_j)}
\end{equation}
In a multilayer perceptron, the outputs of the units in one layer form the inputs to the next layer. The weights of the network are usually computed by training the network using the back-propagation (BP) algorithm.

A multilayer perceptron represents a nested sigmoidal scheme (18), its form for a single output neuron is
\begin{equation}
F(x,\omega)=\varphi(\sum_j\omega_{oj}\varphi(\sum_k\omega_{jk}\varphi(\cdots\varphi(\sum_l\omega_{ll}x_l)\cdots)))
\end{equation}
where $\varphi_j(\cdot)$ is the sigmoidal activation function, $\omega_{oj}$ is the synaptic weight from neuron $j$ in the last hidden layer to the single output neuron $o$, and so on for the other synaptic weights, $x_i$ is the $i$-th element of the input vector $x$. The weight vector $w$ denotes the entire set of synaptic weights ordered by layer, then the neurons in a layer, and then their number in a neuron.

\section{Chosen Sample and Parameters}

Usually, astronomical object classification is based on the properties of spectra, photometry, multiwavelength and so on. In order to check the effectiveness and the efficiency of our provided methods, we classified objects with data from X-ray (ROSAT), optical (USNO-A2.0) and infared (2MASS) bands. By positional cross-correlation of ROSAT, USNO-A2.0 and 2MASS released databases, we obtain the multi-wavelength data. The three catalogs are described in detail as follows: 

The ROSAT All-Sky (RASS) using an imaging X-ray Telescope (Tr$\ddot{u}$mper 1983), are well suited for investigating the X-ray properties of astronomical objects. The RASS Bright Source Catalogue (RBSC) includes 18,811 sources, with a limiting ROSAT PSPC countrate of 0.05 counts s$^{-1}$ in the 0.1-2.4 keV energy band. The typical positional accuracy is 30''. Similarly, the RASS Faint Source Catalogue (RFSC) contains 105,924 sources and represents the faint extension to RBSC. The RBSC and RFSC catalogues contain the ROSAT name, positions in equatorial coordinates, the positional error, the source countrate ($CR$) and error, the background countrate, exposure time, hardness-ratios $HR1$ and $HR2$ and errors, extent ($ext$) and likelihood of extent ($extl$), and likelihood of detection. The two hardness ratios $HR1$ and $HR2$ represent X-ray colors. From the count rate $A$ in the 0.1-0.4 keV energy band and the count rate $B$ in the 0.5-2.0 keV energy band, $HR1$ is given by: $HR1=(B-A)/(B+A)$. $HR2$ is determined from the count rate $C$ in the 0.5-0.9 keV energy band and the count rate $D$ in the 0.9-2.0 keV energy band by: $HR2=(D-C)/(D+C)$. $CR$ is ROSAT total count rate in counts s$^{-1}$. The parameters of $ext$ and $extl$ are source extent in arcsecond and likelihood of source extent in arcsecond, respectively. The amount of $ext$ is specified, by which the source image exceeds the point spread function. The parameters of $ext$ and $extl$ reflect that sources are point sources or extent sources. For example, stars or quasars are point sources; galaxies or galaxy clusters are extent sources. Therefore $ext$ and $extl$ are useful for classification of objects.

The USNO-A2.0 (Monet et al. 1998) is a catalog of 526,280,881 stars over the full sky, compiled in the U.S. Naval Observatory, which contains stars down to about 20 mag over the whole sky. Its astrometric precision is non-uniform, depending on position on Schmidt plates, typically better than 1''. USNO-A2.0 presents right ascension and declination (J2000, epoch of the mean of the blue and red plate) and the blue and red magnitude for each star.

The infrared data is the first large incremental data release from the Two Micron All Sky Survey (2MASS). This release covers 2,483 square degree of northern sky observed from the 2MASS facility at Mt. Hopkins, AZ. The catalogue contains 20.2 million point and 73,980 extended sources, and includes three bands J (1.25 $\mu$m), H (1.65 $\mu$m), and K$_s$ (2.17 $\mu$m) magnitudes.

For supervised methods, the input sample must be tagged with known classes. So the catalogues of known classes of astronomical objects need to be adopted. We choose known AGNs from the catalog of AGN (V\'eron-Cetty \&  V\'eron, 2000), which contains 13214 quasars, 462 BL Lac objects and 4428 active galaxies (of which 1711 are Seyfert 1). Stars include all spectral classes of stars, dwarfs and variable stars, which are adopted from SIMBAD database. Normal galaxies are from Third Reference Catalogue of Bright Galaxies (RC3; de Vaucouleurs et al. 1991).
 
Studying the clustering properties of astronomical objects in a multidimensional parameter space needs catalogue cross-correlation to get multi-wavelength parameters available for all sources. Firstly, within a search radius of 3 times the RBSC and RFSC positional error, we positionally cross-identified the catalogue of USNO-A2.0 with the RBSC and RFSC X-ray sources, and then cross-matched the data from X-ray and optical bands with infared sources in 2MASS first released database within 10 arcsec radius. Secondly, we similarly cross-identified the data from three bands with the catalogues of AGNs, stars and normal galaxies within 5 arcsec radius. Only considering the unique entries, the total sample contains 1656 (29.9\%) AGNs, 3718 (67.0\%) stars, 173 (3.1\%) normal galaxies. 

In the whole process, the obtained data of AGNs, stars and galaxies with catalogue counterparts are divided into four subclasses, (i) unique entries, (ii) multiple entries, (iii) the same entries, (iv) no entries. In detail, unique entries refer to the objects which have only one catalogue entry in the various catalogues, or which have a unique identification in private catalogues. Multiple entries refer to the objects that have more than one catalogue entries in various catalogues. The same entries point to the two or three kinds of objects which have the same catalogue counterparts. No entries show that the objects may not be matched from one or more catalogues, by the reason of the incompleteness of catalogues. In addition, we point out the sample here is obtained by multi-wavelength cross-identification. For positional error, some sources unavoidably match the unrelated or fake sources. In order to keep sources as true as possibly, we only consider the unique entries, cross out the multiple entries, the same entries and no entries. Certainly, knowing which are true sources, we need to compute the probability to assess the validity of identifications of the counterparts from three bands, just like what Mattox et al. 1997, Rutledge et al. 2000 do with cross-association. Owing to the restrictive aim of this work, we don't investigate this respect in detail.

In the paper, the plausibility is based on the optical classification, X-ray characteristics like hardness ratios and extent parameter, and the infrared classification (Stocke et al. 1991; Motch et al. 1998; Pietsch et al. 1998; He et al. 2001). According to the results of the $Einstein$ Medium Sensitivity Survey (EMSS; Stocke et al. 1991), X-ray-to-optical flux ratio, $F_{X}/F_{opt}$, was found to be very different for different classes of X-ray emitters. Motch et al. (1998) stated that, for source classification, the most interesting parameters are flux ratios in various energy bands, including the conventional X-ray hardness ratios, $F_{X}/F_{opt}$ ratios as well as optical colors. They also presented that, although stars and AGNs have similar X-ray colors, their mean X-ray to optical ratios are obviously quite different and they are well separated in the $HR1/2$ vs. $F_{X}/F_{opt}$ diagram. Cataclysmic variables exhibit a large range of X-ray colors and $F_{X}/F_{opt}$ ratios and can be somewhat confused with both AGNs and the most active part of the stellar population. However, the addition of a $B-V$ or $U-B$ optical index would allow to further distinguish between these overlapping population. He et al. (2001) stated that galactic stars usually have bright optical magnitudes and weak X-ray emission, galaxies with fainter optical magnitudes and median X-ray emission, and AGNs with the faintest magnitudes and strongest X-ray emission. In their Figure 1. of $F_{X}/F_{opt}$ vs. $m_{V}$, AGNs and non-AGNs occupy different zones. Pietsch et al. (1998) also used a conservative extent criterion ($extent Likelihood >10''$ and $extent >30''$) as an indicator that the X-ray emission does not originate from a nuclear source. Since the corresponding parameter spaces overlap significantly for different classes of objects, an unambiguous identification based on one band data alone is not possible. In order to classify sources, we consider the data from optical, X-ray and infrared bands. The chosen parameters from different bands are $B-R$ (optical index), $B+2.5log(CR)$ (optical-X-ray index), $CR$, $HR1$ (X-ray index), $HR2$ (X-ray index), $ext$, $extl$, $J-H$ (infrared index), $H-K_s$ (infrared index), $J+2.5log(CR)$ (infrared-X-ray index). Motch et al. (1998) showed that the X-ray to optical flux ratio can be approximate to $log(F_{X}/F_{opt})=log(count rate)+V/2.5-5.63$, assuming an average energy conversion factor of 1 PSPC cts s$^{-1}$ for a 10$^{-11}$ erg cm$^{-2}$ s$^{-1}$ flux in the range of 0.1 to 2.4 keV. So $B+2.5log(CR)$ can be viewed as an X-ray-to-optical flux ratio, similarly, $J+2.5log(CR)$ is an X-ray-to-infrared flux ratio.  

The mean values of parameters for the sample are given in Table 1. Table 1 indicates that some mean values of parameters have rather big scatter. The $B-R$ value of normal galaxies is obviously larger than those of AGNs and stars; the $CR$ value of AGNs is higher than those of stars and normal galaxies. For the mean values of $HR2$, which subdivides the hard range, there are only marginal differences between the individual classes of objects. This applies to the total sample. There is a trend that galaxies seem to have somewhat higher $<$HR2$>$ values than AGNs and stars. AGNs and stars have on the average the lower $HR1$, i.e., they have the softer spectral energy distribution (SED). A significantly harder SED is found for normal galaxies with $<$HR1$>=+0.65$. This is indeed what is expected for this class of objects which exhibit a rather hard intrinsic spectrum caused by thermal bremsstrahlung from a hot ($10^7-10^8$K) plasma(cf. e.g. B$\ddot {o}$hringer 1996). The mean values of $ext$ and $extl$ of normal galaxies is apparently larger than AGNs and stars. Furthermore, those of AGNs are larger than stars. As Table 1 shows, galaxies are not only 0.76 mag in $J-H$, but they also have $H-K$ values, 0.37 mag, redder than stars. Likewise, AGNs are redder than stars, too. We also find that the mean $<$B+2.5log(CR)$>$ and $<$J+2.5log(CR)$>$ values of AGNs are much higher than those of stars and galaxies. This can be explained by the fact that AGNs are strong X-ray emitters.

\begin{table*}[ht]
\begin{center}
\caption{The mean values of parameters for the samples}
\begin{tabular}{rllllll}
\hline
\hline
NO.& parameters&AGNs &stars   &normal galaxies\\
\hline
1 &B-R  &0.41$\pm 0.78$&-1.53$\pm 4.19$&1.42$\pm 1.49$\\
2 &B+2.5log(CR)&13.66$\pm 1.83$&4.18$\pm 5.33$ &7.95$\pm 2.40$\\
3 &CR &0.20$\pm 0.43$&0.12$\pm 0.42$ &0.08$\pm 0.13$\\
4 &HR1&0.09$\pm 0.53$&0.09$\pm 0.53$ &0.65$\pm 0.37$\\
5 &HR2&0.14$\pm 0.41$&-0.02$\pm 0.54$&0.22$\pm 0.48$\\
6 &ext&6.28$\pm 9.52$&4.21$\pm 9.72$ &16.11$\pm 32.12$\\
7 &extl&1.88$\pm 6.62$&1.05$\pm 6.74$ &7.81$\pm 31.15$\\
8 &J-H&0.73$\pm 0.23$&0.24$\pm 1.77$&0.76$\pm 0.17$\\
9 &H-K$_s$&0.76$\pm 0.27$&0.09$\pm 0.11$&0.37$\pm 0.19$\\
10 &J+2.5log(CR)&12.80$\pm 1.27$&4.33$\pm 1.8$&9.75$\pm 1.54$\\
\hline
\end{tabular}
\bigskip
\end{center}
\end{table*}

In order to see the difference of astronomical objects, we plot the statistical histograms of objects similar to the method used in Voges et al. (1999). In Fig.1 we present the distributions of ten parameters of AGNs and S\&G. Here S\&G is short for stars and normal galaxies. The horizontal axes are labeled by all kinds of parameters and the vertical axes are labeled by the number of sources. From the distributions of $B-R$, $B+2.5logCR$, $J-H$, $H-K$, $J+2.5logCR$, it is obvious that AGNs are different from stars and galaxies. While for the distributions of $CR$, $HR1$, $HR2$, $ext$, $extl$, AGNs overlap seriously with stars and galaxies. In other words, $B+2.5log(CR)$ and $J+2.5log(CR)$ are the most important attributes to be used for classification. $B-R$, $J-H$ and $H-K$ are more important. The others contribute little. To determine the best combination of parameters for AGNs, stars and galaxies discrimination, we have 2-dimensional, 5-dimensional and ten-dimensional spaces to probe. Hereafter dimensional is short for D. The goal of the paper is to verify the discriminant of learning vector quantization (LVQ), support vector machines (SVM) and single-layer perceptron (SLP). In the following section, we explore this respect.

\section{Results and Discussion}

\subsection{Results}

Since $B+2.5log(CR)$ and $J+2.5log(CR)$ may be used as important features, we select $B+2.5log(CR)=11.8$ and $J+2.5log(CR)=10.5$ as classification criterion to discriminate stars and galaxies from AGNs. If $B+2.5log(CR)\ge 11.8$, the objects belong to AGNs, otherwise, to stars and galaxies. Similarly, if $J+2.5log(CR)\ge 10.5$, the objects belong to AGNs, otherwise, to stars and galaxies. The situation is divided into three: only considering $B+2.5log(CR)=11.8$, only considering $J+2.5log(CR)=10.5$ and considering both criterions. The results of classification are presented in Table 2. The whole accuracy for each situation is 94.0\%, 96.5\% and 94.9\%, respectively. Evidently, the results are comparable for three situations.

\begin{table*}[ht]
\begin{center}
\caption{Result of Classification for three situations}
\begin{tabular}{rllllll}
\hline
\hline
&1&&2&&3\\
\hline
classified$\downarrow$known$\to$& AGNs &S\&G& AGNs &S\&G& AGNs &S\&G\\
\hline
        AGNs & 1475  & 151&1574& 112&1598&226 \\
        S\&G & 181  &3740& 82  &3779&58&3665\\
\hline
 accuracy   & 89.0\% & 96.0\%& 95.0\%&97.0\%&96.5\%&94.0\%\\
\hline
Total accuracy&94.0\% &&96.5\% &&94.9\%\\
\hline
\end{tabular}
\bigskip
\end{center}
\end{table*}

In order to understand which parameter combination is best, we explore the bidimensional space composed of $B+2.5log(CR)$ and $J+2.5log(CR)$, 5D space composed of $B+2.5log(CR)$, $B-R$, $J-H$, $H-K$ and $J+2.5log(CR)$, and 10D space composed of the ten parameters. Randomly dividing the sample into two parts, one as the training set and another as the test set, we use different methods to train the training set and get different classifiers. Then with the test set, we check how the classifiers are when applied for classification. If good, the classifiers can be used for predicting the unknown sources. Firstly, we apply learning vector quantization (LVQ) to classify AGNs from stars and normal galaxies. The results are given in Table 3. In 2D, 5D and 10D spaces, the total accuracy is 97.66\%, 97.69\% and 97.80\%, respectively. Secondly, we employ support vector machines (SVM) in different spaces. The computed results are shown in Table 4. As Table 4 shows, the total accuracy amounts to 75.52\%, 98.09\% and 98.31\%, respectively. Comparing the results with those by single-layer perceptron (SLP), we give the result in Table 5. We train a perceptron with two input neurons, one output neuron and no hidden neurons for 1000 epochs. In 2D, 5D, 10D spaces, the total accuracy adds up to 97.69\%, 98.09\%, 98.05\%, respectively. 

From the results by LVQ, SVM, SLP, it is obvious that the performances of them are comparable. In low dimensional spaces, LVQ and SLP is better than SVM, but in high dimensional spaces, SVM shows its superiority. Just comparing LVQ and SLP, SLP is better. Considering enough attributes for classification, the automated methods are superior to the simple physical cutoff. Moreover, the high accuracy suggests that the useful features for classification can be extracted by the method of histogram, i.e. the histogram method may be used as the applicable feature selection technique.

\begin{table*}[ht]
\begin{center}
\caption{Result of Classification by LVQ}
\begin{tabular}{rllllll}
\hline
\hline
&2D space&&5D space&&10D space\\
\hline
classified$\downarrow$known$\to$& AGNs &S\&G& AGNs &S\&G& AGNs &S\&G\\
\hline
        AGNs &824&61&828&64&828&61\\
        S\&G &4&1885&0&1882&0&1885\\
\hline
               accuracy&99.52\%&96.87\%&100.0\%&96.71\%&100.0\% &96.87\%\\
\hline
Total accuracy&97.66\% &&97.69\% &&97.80\%\\
\hline
\end{tabular}
\bigskip
\end{center}
\end{table*}

\begin{table*}[ht]
\begin{center}
\caption{Result of Classification by SVM}
\begin{tabular}{rllllll}
\hline
\hline
&2D space&&5D space&&10D space\\
\hline
classified$\downarrow$known$\to$& AGNs &S\&G& AGNs &S\&G& AGNs &S\&G\\
\hline
        AGNs &150&1&782&7&818&37\\
        S\&G &678&1945&46&1939&10&1909\\
\hline
               accuracy&18.12\%&99.95\%&94.44\%&99.64\% & 98.79\%  & 98.10\% \\
\hline
Total accuracy&75.52\% &&98.09\% &&98.31\%\\
\hline
\end{tabular}
\bigskip
\end{center}
\end{table*}

\begin{table*}[ht]
\begin{center}
\caption{Result of Classification by SLP}
\begin{tabular}{rllllll}
\hline
\hline
&2D space&&5D space&&10D space\\
\hline
classified$\downarrow$known$\to$& AGNs &S\&G& AGNs &S\&G& AGNs &S\&G\\
\hline
        AGNs &826&62&826&51&825&51\\
        S\&G &2&1884&2&1895&3&1895\\
\hline
   accuracy&99.76\%&96.81\%&99.76\%&97.38\% & 99.64\%  & 97.38\% \\
\hline
Total accuracy&97.69\% &&98.09\% &&98.05\%\\
\hline
\end{tabular}
\bigskip
\end{center}
\end{table*}

\subsection{Discussion}
Table 2 shows that the efficiency of Classification is rather high, more than 90\% when only considering the important features. Apparently, it is simple and applicable to choose a few good features for classification. But compared to the results by the automated algorithms, such a method is a little inefficient. After all, the method is limited by itself for it can't avoid losing information only with a few features. What's more, sometimes it is very difficult to find such good features. Only depending on other tools, such as principal component analysis (Folkes et al. 1996, Zhang et al. 2003), we can find the principal features. If the number of principal components is more than 3, it is not appliable to use simple cutoff for the difficulty of visualization. As a result, it is better to apply automatic approaches under such situations.

For LVQ and SLP, as shown by Tables 3 and 5, the results are rarely affected by the number of space dimension when the space owns the important features. But for SVM, in contrast, the result of Table 4 is closely connected with the number of space dimension even including the important features. Moreover, the more parameters considered, the higher the accuracy is. For low dimensional spaces, LVQ and SLP are better. While for high dimensional spaces, SVM shows its superiority. Moreover, the statistics listed in Tables 3-5 give a view of how well the algorithms did in classifying AGN and non-AGN objects. These statistics tell us how effective a given method is at correctly identifying a true AGN as an AGN or a true non-AGN as a non-AGN. In other words, how often does the method misidentify objects? If the number of AGN objects identified as non-AGNs were zero, the classified accuracy of AGNs is 100\%. Conversely, if the number of non-AGNs identified as AGNs were zero, the classified accuracy of stars and normal galaxies is 100\%. The generally lower values of the classified accuracy of AGNs compared to those of stars and normal galaxies may be a result of the smaller sample size for AGNs (1656 vs. 3891). This suggests that it would be useful to run these tests again with a larger sample base for the methods examined here. Given our results for the methods presented here, we are encouraged that distinguishing between a number different types of objects should be possible. For such a project, a larger number of samples of each type of object would be necessary to have an adequate ability to distinguish between the classes. Comparing the computed results, we conclude that LVQ, SVM and SLP are effective methods to classify sources with multi-wavelength data. With the data from three bands, we can classify AGNs from stars and normal galaxies effectively by LVQ, SLP or SVM. This also indicates that the chosen parameters are such good feature vectors to separate AGNs from stars and normal galaxies. We believe the performance will increase if the data are complete or the quality and quantity of data improves. Moreover, these methods can be used to preselect AGNs from large numbers of sources in large surveys avoiding wasting time and energy, when studying AGNs or cosmology. The three supervised learning methods we investigated here gave comparable results in a number of situations. Generally, the more features considered, the better results SVM gave; however,the results of LVQ and SLP were considerable with different number of attributes. Also, the different methods, while giving different quality results in a number of cases, were comparable for most of the samples we examined. However, our results suggest that the parameters we choose did not adequately pick out characteristics of the objects in all cases. Other parameters added from more bands that effectively summarize the features of sources, such as from radio band, appear to do better (Krautter et al. 1999). Thus we can improve the classified accuracy of AGNs or stars and normal galaxies, even classify different types of AGNs. Moreover, these methods can be used for other types of data, such as spectral data and photometric data. We believe that it would be beneficial to have more extensive comparisons between different methods. Only then can we take some of the magic out of determining what parameters to choose and know which method to use better in different cases.

The performances of LVQ and SLP are different from that of SVM, which arises from different methods based on different theories. SVM embodies the Structural Risk Minimization (SRM) principle, which is superior to Empirical Risk Minimization (ERM) principle that conventional neural networks employ. Most neural networks including LVQ and SLP are designed to find a separating hyperplane. This is not necessarily optimal. In fact many neural networks start with a random line and move it, until all training points are on the right side of the line. This inevitably leaves training points very close to the line in a non-optimal way. However, in SVM, a large margin classifier, i.e. a line approaching the optimal is sought. As a result, SVM shows better performance than LVQ and SLP in the high dimensional space.    

\section{Conclusion}

Sources classification depends on the quality and amount of real-time data and on the algorithm used to extract generalized mappings. Availability of the high-resolution multi-wavelength data constantly increases. The best possible use of this observational information requires efficient processing and generalization of high-dimensional input data. Moreover, good feature selection techniques, as well as good data mining methods, are in great demand. A very promising algorithm that combines the power of the best nonlinear techniques and tolerance to very high-dimensional data is support vector machines (SVM). In this work we have used histogram as the feature selection technique and applied LVQ, SLP and SVM to multi-wavelength astronomy to classify AGNs from stars and normal galaxies. We conclude that the features selected by histogram are applicable and the performance of SVM models can be comparable to or be superior to that of the NN-based models in the high dimensional space. The advantages of the SVM-based techniques are expected to be much more pronounced in future large multi-wavelength survey, which will incorporate many types of high-dimensional, multi-wavelength input data once real-time availability of this information becomes technologically feasible. All these methods can be used for astronomical object classification, data mining and preselecting AGN candidates for large survey, such as the Large Sky Area Multi-Object Fiber Spectroscopic Telescope (LAMOST). Various data, incuding morphology, photometry, spectral data and so on, can be applied to train the methods and obtain classifiers to classify astronomical objects or preselect intresting objects. When lacking training sets, we may explore some unsupervised methods or outlier finding algorithms to find unusual, rare, or even new types of objects and phenomena. In addition, with the development of the Virtual Observatory, these methods will be part of the toolkits of the International Virtual Observatory.

\begin{acknowledgements}
We are very grateful to anonymous referee for his important comments and suggestions. We would like to thank LAMOST staff for sincere help. This research has made use of the SIMBAD database, operated at CDS, Strasbourg, France. Simultaneously, this paper has also made use of data products from the Two Micron All Sky Survey, which is a joint project of the University of Massachusetts and the Infrared Processing and Analysis Center/California Institute of Technology, funded by the National Aeronautics and Space Administration and the National Science Foundation. This research is supported by National Natural Science Foundation of China under Grant No.10273011.
\end{acknowledgements}

\begin{figure}
\input epsf
\epsfverbosetrue
\epsfxsize 10cm
\epsfbox{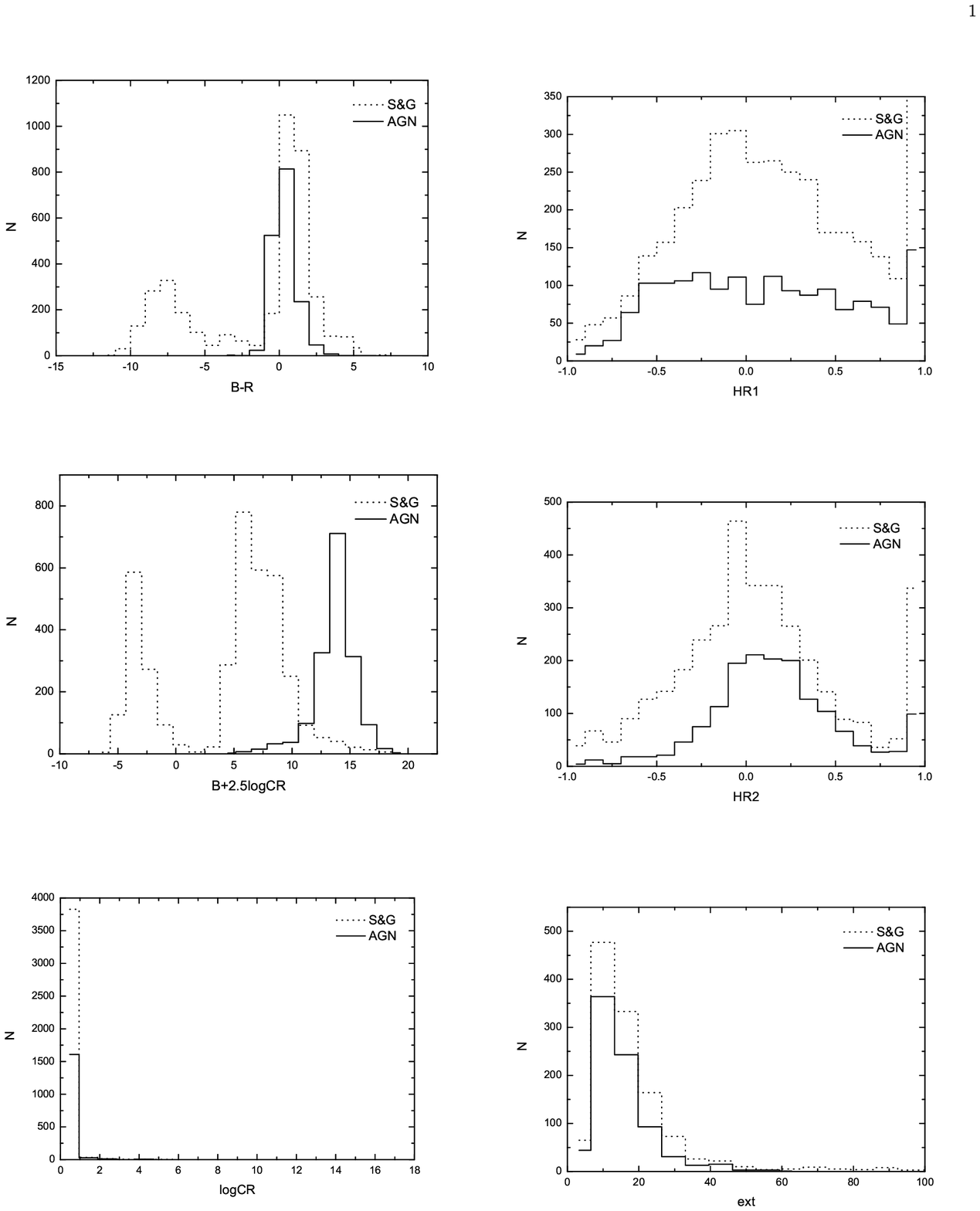}
\end{figure}

\newpage

\begin{figure}
\input epsf
\epsfverbosetrue
\epsfxsize 10cm
\epsfbox{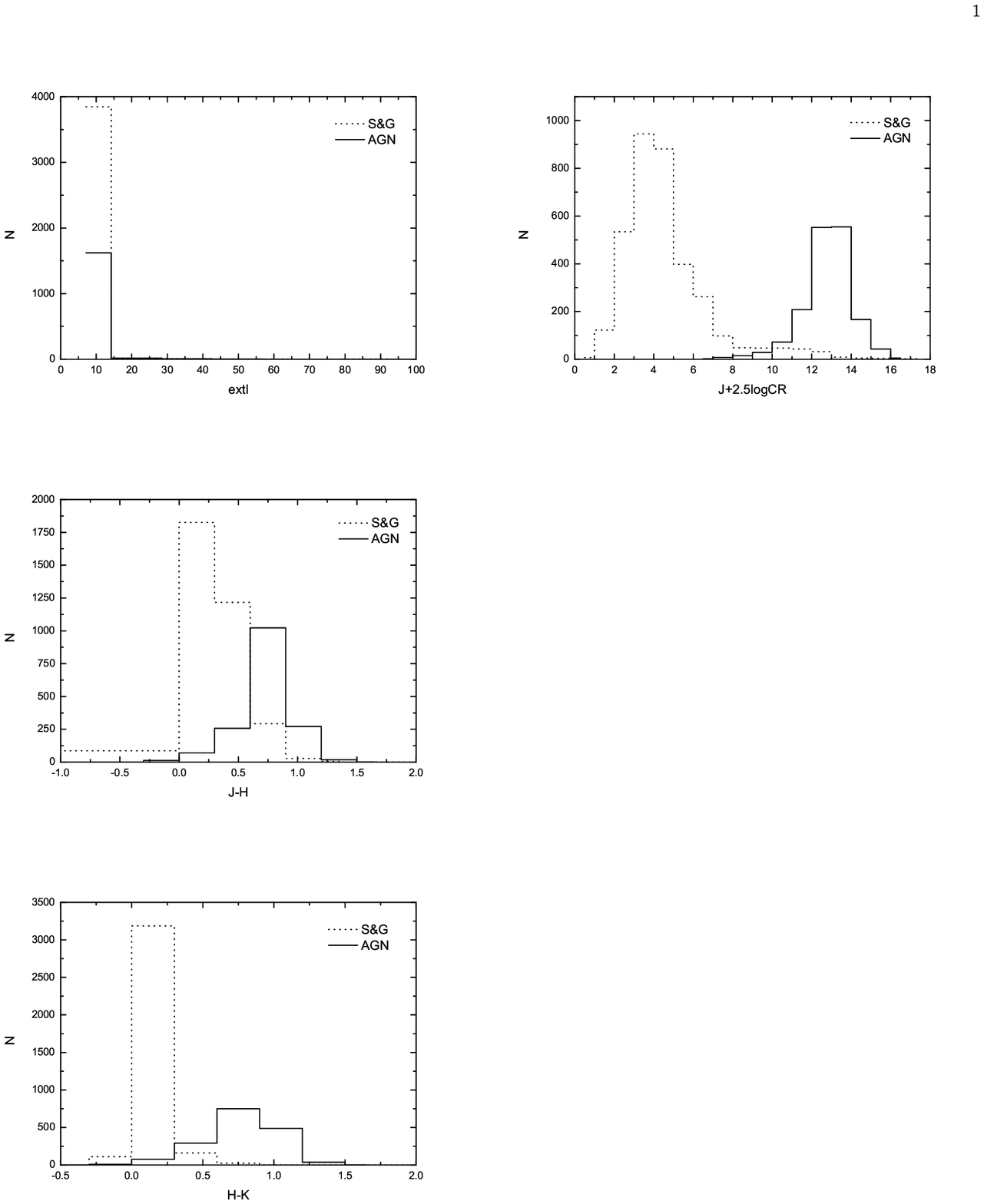}
\caption{Ten histograms summarizing some results of the
analysis of RBSC and RFSC sources for 1656 AGNs (solid line) and
3891 S\&G (dotted line).}
\end{figure}

\end{document}